\documentclass[a4paper,11pt]{article}
\pdfoutput=1

\usepackage{jcappub}
\usepackage{amsmath}
\usepackage{amssymb}
\usepackage{subcaption}
\usepackage{graphicx}
\usepackage{booktabs}
\usepackage{geometry}
\usepackage{natbib}
\usepackage{comment}
\usepackage[colorlinks=true, linkcolor=blue, citecolor=blue, urlcolor=blue]{hyperref}

\title{\boldmath Refining the Greisen Profile for Low-Energy Cosmic Gamma-Rays: Quantifying Deviations Across Altitudes and Zenith Angles}

\author[a,c,e]{Constanza Valdivieso,}
\author[a]{Bárbara Gutiérrez,}
\author[a,c]{Nicolás Viaux M.}
\author[a]{Sebastián Mendizabal,}
\author[b,d]{Raquel Pezoa R.,}
\author[a,d]{Sebastian Tapia}

\affiliation[a]{Departamento de Física, Universidad Técnica Federico Santa María, Valparaíso, Chile}
\affiliation[b]{Departamento de Informática, Universidad Técnica Federico Santa María, Valparaíso, Chile}
\affiliation[c]{Millennium Institute for Subatomic physics at high energy frontier (SAPHIR), Santiago, Chile}
\affiliation[d]{Centro Cientifico Tecnólogico de Valparaíso (CCTVal), Universidad Técnica Federico Santa María, Valparaíso, Chile}
\affiliation[e]{Instituto de Física, Pontificia Universidad Católica de Valparaíso,
Avenida Universidad 331, Curauma, Valparaíso, Chile}
\emailAdd{nicolas.viaux@usm.cl}

\abstract{This study refines the Greisen formalism by comparing the classical Greisen profile and a modified Greisen profile, which incorporates an empirical correction to the shower age parameter \( s \), both with zenith-angle dependence, aiming to better describe low-energy cascades, against CORSIKA simulations of cosmic gamma-ray showers (20–800 GeV). Fittings across altitudes of 5000–5900 m and zenith angles of 0$^\circ$–40$^\circ$ quantify deviations in particle numbers, showing the modified profile yields deviations below 4.7\%, compared to up to 12.5\% for the classical profile. These improvements address low-energy ionization losses, atmospheric density variations, and zenith-angle effects, enhancing accuracy for high-altitude observatories like HAWC and the proposed CONDOR array. The modified profile offers a computationally efficient alternative, providing precise particle number predictions to advance gamma-ray astrophysics and cosmic-ray research.}

\keywords{Cosmic Rays, Air-Showers, Greisen Function, CONDOR Observatory}

\begin{document}
\maketitle
\flushbottom

\section{Introduction}
\label{sec:intro}

Extensive air showers (EAS), cascades of secondary particles initiated when high-energy cosmic rays interact with atmospheric nuclei, are a cornerstone of astroparticle physics. They encode essential information about the primary particle’s energy, composition, and interaction mechanisms, offering insights into the origin of cosmic rays and high-energy astrophysical processes \cite{Aab2014, Abbasi2018}. In the case of cosmic gamma rays, showers are dominated by electromagnetic processes such as pair production and bremsstrahlung, producing secondary electrons and positrons \cite{Rossi1941}. Unlike hadronic showers initiated by protons or heavier nuclei, which involve complex nuclear interactions, gamma-ray–induced showers are simpler to model, making them an ideal benchmark for testing analytical frameworks such as the Greisen formalism \cite{Greisen1956}. Their longitudinal development, defined by the number of particles as a function of atmospheric depth, provides a clean probe of the primary gamma-ray energy and the underlying atmospheric interaction processes \cite{Gaisser1990}.\\

A key analytical description of electromagnetic cascades is the Greisen formalism, based on the cascade theory of Rossi and Greisen. This model provides a compact expression for the particle number $N(t)$ as a function of radiation length $t$, assuming energy regimes where radiative processes dominate over ionization losses. While the Greisen profile has been applied at ultra-high energies ($E \gtrsim 10^{17}$ eV), its accuracy becomes limited at lower energies due to increased ionization losses and reduced particle multiplicity \cite{Lipari2009}. Recent efforts have introduced modifications to include zenith-angle dependence, but systematic validation in the low-energy regime remains scarce.\\

High-altitude observatories, such as the HAWC array at 4100 m \cite{Abeysekara2017} and the proposed CONDOR array at 5300 m \cite{Arratia2023}, are particularly sensitive to low-energy gamma-ray showers. At elevations of 5000–5900 m, the reduced atmospheric density increases the effective radiation length and alters cascade development, while zenith-angle variations change the atmospheric path length and shift the shower maximum. These factors can introduce discrepancies between the Greisen profile and observed showers, underscoring the need for careful validation.\\

Monte Carlo simulations such as CORSIKA (COsmic Ray SImulations for KAscade) \cite{Heck1998} provide detailed reconstructions of shower profiles and are the standard for precision studies, but they remain computationally intensive. Analytical models, when validated, offer a complementary and efficient alternative for rapid interpretation of low-statistics datasets or real-time applications in high-altitude observatories \cite{AlvarezMuniz2012, PierreAugerCollaboration2021}.\\

This study addresses three gaps in air shower modeling for low-energy cosmic gamma-rays:
\begin{itemize}
    \item The lack of systematic validation of the Greisen formalism for low-energy cosmic gamma-ray showers (20--800 GeV) where in this energy range fewer particles arrive to the ground from a cosmic-ray hitting the atmosphere, due to reduced particle multiplicity and increased ionization losses.
    \item The need to quantify deviations in particle numbers for both the classical and modified Greisen profiles compared to CORSIKA simulations across zenith angles (0$^\circ$--40$^\circ$) and high altitudes (5000--5900 m), where atmospheric and geometric effects are significant.
    \item The reconciliation of analytical models with Monte Carlo simulations to address uncertainties arising from atmospheric density variations and simulation-specific assumptions in electromagnetic shower modeling.
\end{itemize}

By fitting the classical and modified Greisen profiles to CORSIKA simulations across altitudes and zenith angles, we quantify deviations in particle numbers using a parameter \(\alpha\) to measure deviation from CORSIKA. This work refines the Greisen formalism’s applicability for next-generation high-altitude observatories, enhancing its utility as a computationally efficient tool for gamma-ray astrophysics and cosmic-ray studies.\\

\section{Modified Greisen Profile for Low-Energy Cosmic Gamma-Ray Showers}
\label{sec:greisen_function}

The Greisen formalism is a foundational analytical approach for describing the longitudinal development of EAS initiated by low-energy cosmic gamma-rays in the energy range 20–800 GeV. It provides an approximate solution to the cascade equations by modeling the balance between pair production, which multiplies particles at high energies and bremsstrahlung, which sustains the cascade by generating additional  photons. Through these approximations, the formalism predicts the number of particles as a function of atmospheric depth, offering a computationally efficient alternative to Monte Carlo simulations.

The classical Greisen profile is derived from cascade theory under Approximation A, originally introduced by Rossi and Greisen \cite{Rossi1941, Greisen1956} and later reviewed in detail in \cite{Lipari2009}, considers only radiative processes like bremsstrahlung and pair production, and Approximation B, which incorporates ionization losses (energy loss through collisions with atmospheric molecules) and Coulomb scattering (deflection of charged particles by atomic fields) \cite{Rossi1941}. The profile expresses the number of particles \( N(t) \) as a function of the radiation length \( t = (X - X_1)/X_0 \), where: \( X \) is the atmospheric depth, the integrated mass of air traversed by the shower, measured in g$\cdot$cm$^{-2}$, representing the cumulative interaction potential, calculated as:
  \begin{equation}
  X(h, \theta) = \int_h^\infty \rho(z) \sec\theta \, dz,
  \label{eq:atm_depth}
  \end{equation}
  where \(\rho(z)\) is the atmospheric density at height \( z \), and \(\sec\theta = 1/\cos\theta\) accounts for the slant path through the atmosphere due to the zenith angle \(\theta\) \cite{Gaisser1990}. This integral quantifies the mass of air encountered by the shower, critical for modeling the interaction probability and cascade development. \( X_1 \) is the depth of the first interaction, typically near the top of the atmosphere for gamma-rays. \( X_0 \simeq 37 \) g cm$^{-2}$ is the radiation length in air, the average distance over which a high-energy electron loses all but 1/e of its energy via bremsstrahlung.

The Greisen profile is an approximated analytical solution that describe the shower development:
\begin{align}
    \lambda_1(s)=\frac{1}{N(t)}\frac{dN(t)}{dt}.
\end{align}
which gives a the number of particles:
\begin{align}
N(t)=N_0 e^{\int\lambda_1(s)dt},
\end{align}
where the relation between the parameters $t$ and $s$ is given by:
\begin{align}
    t=-\frac{1}{\lambda'_1(s)}[\beta-1/s]\approx-\frac{\beta}{\lambda'_1(s)},
\end{align}
which gives the condition $t_{max}=\beta$ and $\lambda'(s_{max})=-1.$ The first approximation for the Greisen profile $\lambda(s)$ can be then obtain from:
\begin{align}
    \frac{d\lambda_1(s)}{ds}\approx\frac{1}{2}\left(1-\frac{3}{s}\right),
\end{align}
which gives the well known profile:
\begin{align}
    \lambda_1(s)=\frac{1}{2}(s-3\ln s)+c.
\end{align}

The constant $c=-1/2$ can be chosen to agree with the solution of the full cascade evolution equation for a initial value of $t_i=0$. We are freely to choose different initial condition and keep a small fixing parameter, in order to include more realistic atmosphere description we use:
$$c=-1/2-3/2\ln (c's),$$ 
with $c'=0.99745$, so that a new modified profile can be written as:
\begin{align}
    N(t) = N_0 \exp\left[ t\left(1 - \frac{3}{2} \ln (c's) \right) \right],
    \label{eq:greisen_fct}
\end{align}

The classical Greisen profile, fitted to CORSIKA simulations using a parameter \(\alpha\) to measure deviations, is given by:
\begin{align}
    N(t,\theta) = \alpha \frac{0.326}{\sqrt{\beta \cos{\theta}}} \exp\left[ \frac{t}{\cos{\theta}} \left(1 - \frac{3}{2} \ln (c's) \right) \right],
    \label{eq:greisen_fct}
\end{align}
where: \(\beta = \ln(E_0 / E_{\text{cut}})\) is a dimensionless parameter quantifying the shower’s energy scale relative to the critical energy, with \( E_0 \) being the primary gamma-ray energy (20--800 GeV) and \( E_{\text{cut}} \simeq 87 \) MeV the critical energy, where ionization losses equal radiative losses for electrons in air \cite{Gaisser1990}. \(\theta\) is the zenith angle, and the term \(\cos\theta\) adjusts for the increased atmospheric path length in non-vertical showers, as described by Equation \ref{eq:atm_depth}. The shower age parameter \( s \), defined as:
\begin{align}
    s = \frac{3t/\cos{\theta}}{t/\cos{\theta} + 2\beta},
    \label{eq:shower_age}
\end{align}
characterizes the evolutionary stage of the shower: \( s = 0 \) at initiation (first interaction), \( s = 1 \) at the shower maximum (peak particle number), and \( s > 1 \) in the decaying phase where absorption dominates \cite{Lipari2009}. The coefficient 0.326 is a normalization constant derived from asymptotic approximations in electromagnetic cascade theory, ensuring accurate particle counts at the shower maximum.\\

The Greisen formalism faces significant challenges when applied to low-energy cosmic gamma-ray showers at high altitudes (5000--5900 m):
\begin{itemize}

    \item \textit{Atmospheric Variations}: At high altitudes, the air density is approximately 50\% of sea-level density, shortening the radiation length and shifting the shower maximum closer to the observer, which alters the cascade dynamics and requires altitude-specific adjustments \cite{Weekes2002}.
    \item \textit{Zenith-Angle Dependencies}: For zenith angles up to 40$^\circ$, the effective atmospheric depth increases by a factor of \( 1/\cos\theta \) (approximately 1.3 at 40$^\circ$), enhancing multiple Coulomb scattering and photon absorption, which the classical profile may not fully capture \cite{Gaisser1990, AlvarezMuniz2012}.
    \item \textit{Model Simplifications}: Assumptions such as a constant radiation length and negligible low-energy losses overlook atmospheric density gradients, temperature variations, and molecular composition differences (e.g., nitrogen, oxygen ratios), introducing potential inaccuracies \cite{Lipari2009}.
\end{itemize}

In this work we propose a  modified Greisen profile , where we introduce an empirical adjustment to the shower age term to better capture low-energy effects and zenith-angle variations:
\begin{align}
    N(t,\theta) = \alpha \frac{0.326}{\sqrt{\beta \cos{\theta}}} \exp\left[ \frac{t}{\cos{\theta}} \left(1 - \frac{3}{2} \ln (c'\cdot s)) \right) \right],
    \label{eq:modified_greisen_fct}
\end{align}


Validating the Greisen and the modified Greisen profiles against Monte Carlo simulations ensures accurate particle number predictions, which is essential for interpreting low-energy gamma-ray showers at high-altitude observatories \cite{Abeysekara2017, Arratia2023}. 

\section{CORSIKA Simulation Setup}
\label{sec:corsika_setup}

CORSIKA is a premier Monte Carlo simulation package widely used in astroparticle physics to model the initiation and propagation of EAS induced by cosmic rays, including gamma-rays, protons, and heavy nuclei \cite{Heck1998}. Monte Carlo methods simulate stochastic processes by generating random samples to approximate complex physical systems, tracking individual particle trajectories and interactions through the atmosphere with high fidelity. CORSIKA integrates sophisticated interaction models, including high-energy hadronic models (e.g., QGSJET for nuclear interactions, EPOS for high-energy collisions) and electromagnetic models (e.g., EGS4 for photon and electron processes), making it a standard tool for validating analytical models like the Greisen profile and interpreting observational data from ground-based observatories. Its ability to model the full cascade, including secondary particle production, scattering, and absorption, provides a robust benchmark for assessing the accuracy of analytical predictions.

Simulations were conducted at three representative high altitudes 5000 m, 5300 m, 5600 m and 5900 m corresponding to atmospheric pressures of approximately 540--480 hPa, which reduce air density by ~50\% compared to sea-level conditions (1013 hPa). The atmospheric density at height \( h \) is modeled as:
\begin{equation}
\rho(h) = \rho_0 \exp\left(-\frac{h}{H}\right),
\label{eq:density}
\end{equation}
where \(\rho_0 \approx 1.225 \, \text{kg/m}^3\) is the sea-level density, and \( H \approx 8 \, \text{km} \) is the scale height \cite{Gaisser1990}. This lower density shortens the radiation length, shifting the shower maximum closer to the ground and altering cascade dynamics, which is particularly relevant for high-altitude observatories like the Condor Array \cite{Arratia2023}. Zenith angles were varied from 0$^\circ$ (vertical incidence) to 40$^\circ$ in 2$^\circ$ increments, totaling 21 angular configurations per energy and altitude. The effective atmospheric depth for oblique showers is given by:
\begin{equation}
X_{\text{slant}}(\theta) = \frac{X}{\cos\theta},
\label{eq:slant_depth}
\end{equation}
where \( X \) is the vertical atmospheric depth (Equation \ref{eq:atm_depth}), and \( 1/\cos\theta \) (e.g., ~1.3 at 40$^\circ$) accounts for the increased path length, leading to elongated cascades with enhanced multiple scattering and photon absorption \cite{AlvarezMuniz2012}.

To ensure statistical robustness, we performed 1000 independent simulation runs for each combination of primary energy (12 values: 20, 30, 100, 130, 150, 200, 300, 400, 500, 600, 700, 800 GeV), altitude, and zenith angle, using unique random seeds to sample fluctuations in interaction probabilities. This resulted in a comprehensive dataset of 1,008,000 simulations (12 energies × 4 altitudes × 21 angles × 1000 runs), executed on a local computational cluster to manage the significant computational demand.

CORSIKA’s electromagnetic interaction module, based on the EGS4 framework, was configured to accurately simulate key processes: bremsstrahlung, pair production, and ionization. The cross-sections for these processes were derived from quantum electrodynamics, ensuring high precision in modeling particle interactions \cite{Heck1998}. The atmospheric model was customized using the US Standard Atmosphere, adjusted for high-altitude conditions to reflect realistic density gradients (Equation ~\ref{eq:density}) and composition (78\% Nitrogen, 21\% Oxygen, 1\% Argon). A cutoff energy of 1 MeV was set for electrons and photons to balance computational efficiency with accuracy, ensuring all relevant particles were tracked until they were absorbed or reached ground level.

\section{Results: Quantifying Deviations in the Greisen Profile}
\label{sec:results}

\begin{figure}[ht]
\centering
\begin{subfigure}[b]{0.47\textwidth}
    \centering
    \includegraphics[width=\textwidth]{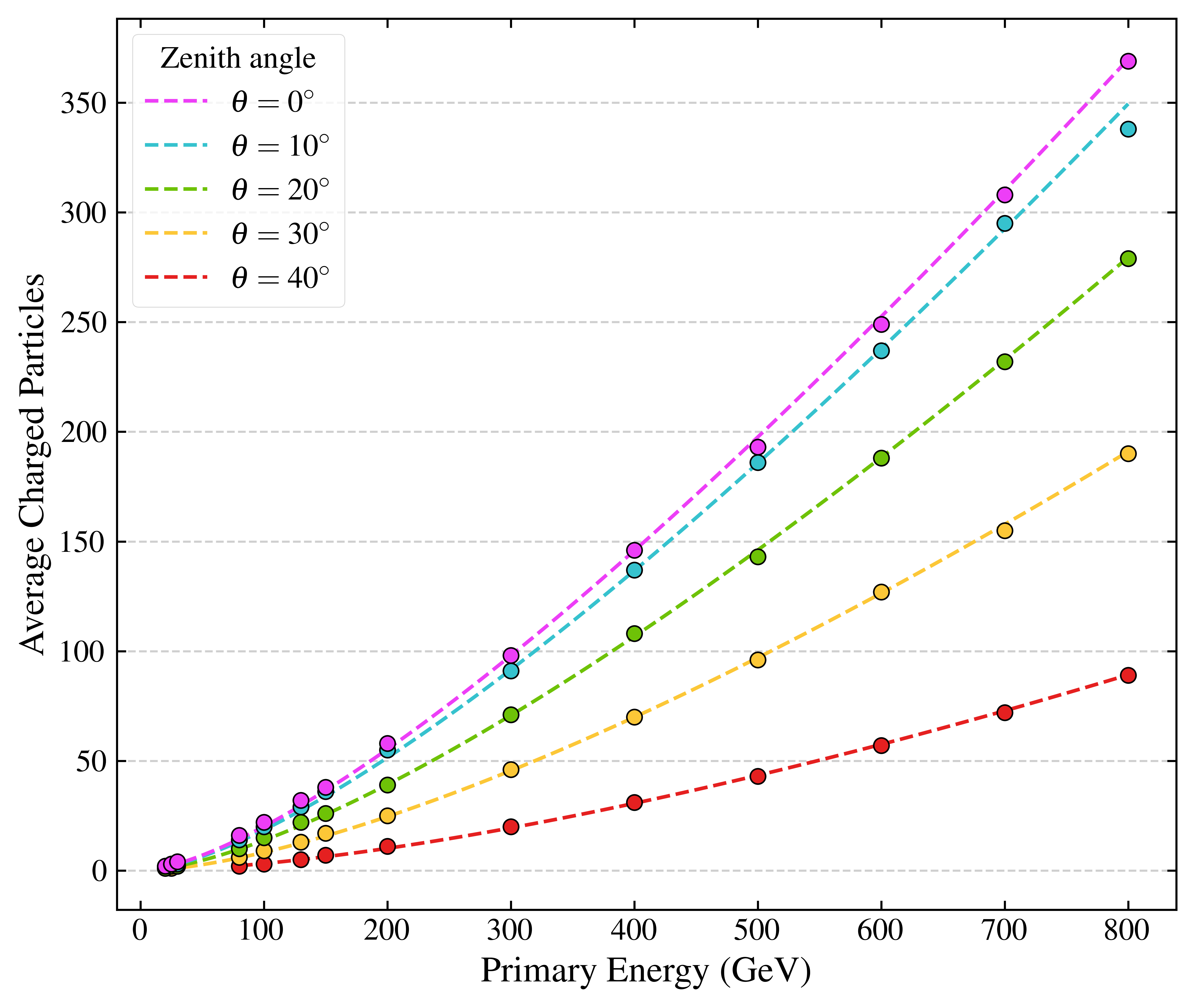}
    \caption{5000 m}
    \label{subfig:profile_1-5000}
\end{subfigure}
\hfill
\begin{subfigure}[b]{0.47\textwidth}
    \centering
    \includegraphics[width=\textwidth]{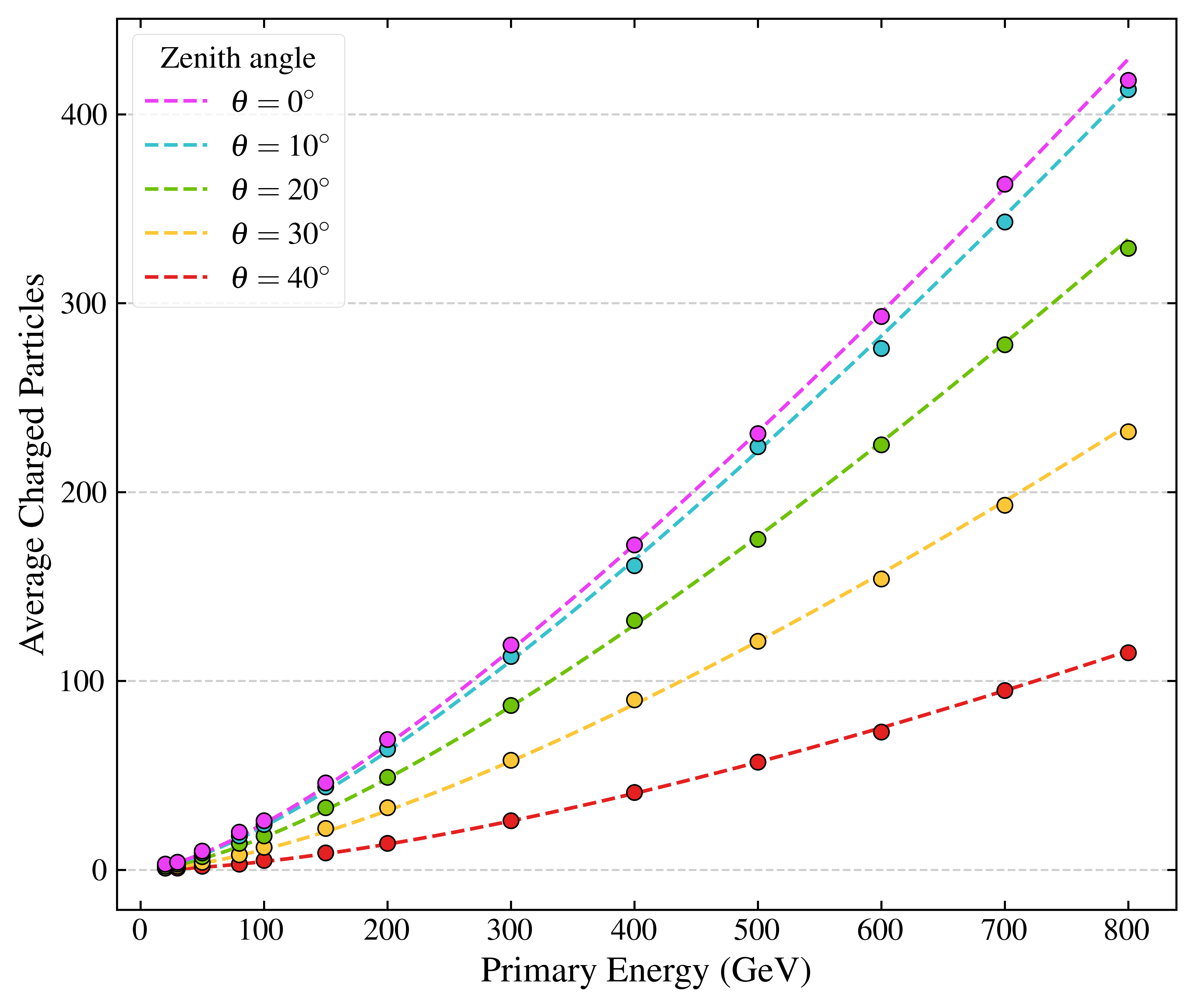}
    \caption{5300 m}
    \label{subfig:profile_1-5300}
\end{subfigure}

\vspace{1em}

\begin{subfigure}[b]{0.47\textwidth}
    \centering
    \includegraphics[width=\textwidth]{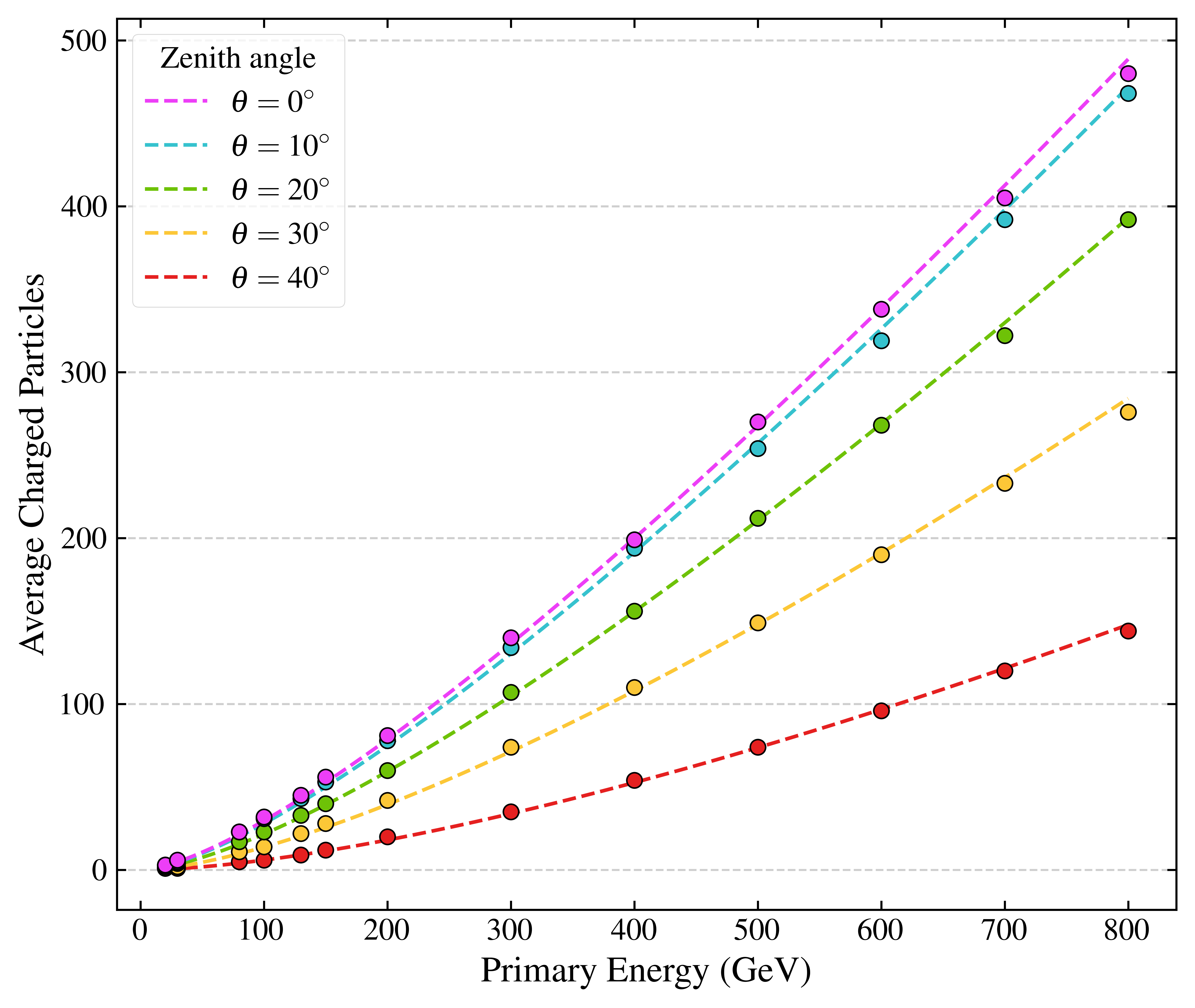}
    \caption{5600 m}
    \label{subfig:profile_1-5600}
\end{subfigure}
\hfill
\begin{subfigure}[b]{0.47\textwidth}
    \centering
    \includegraphics[width=\textwidth]{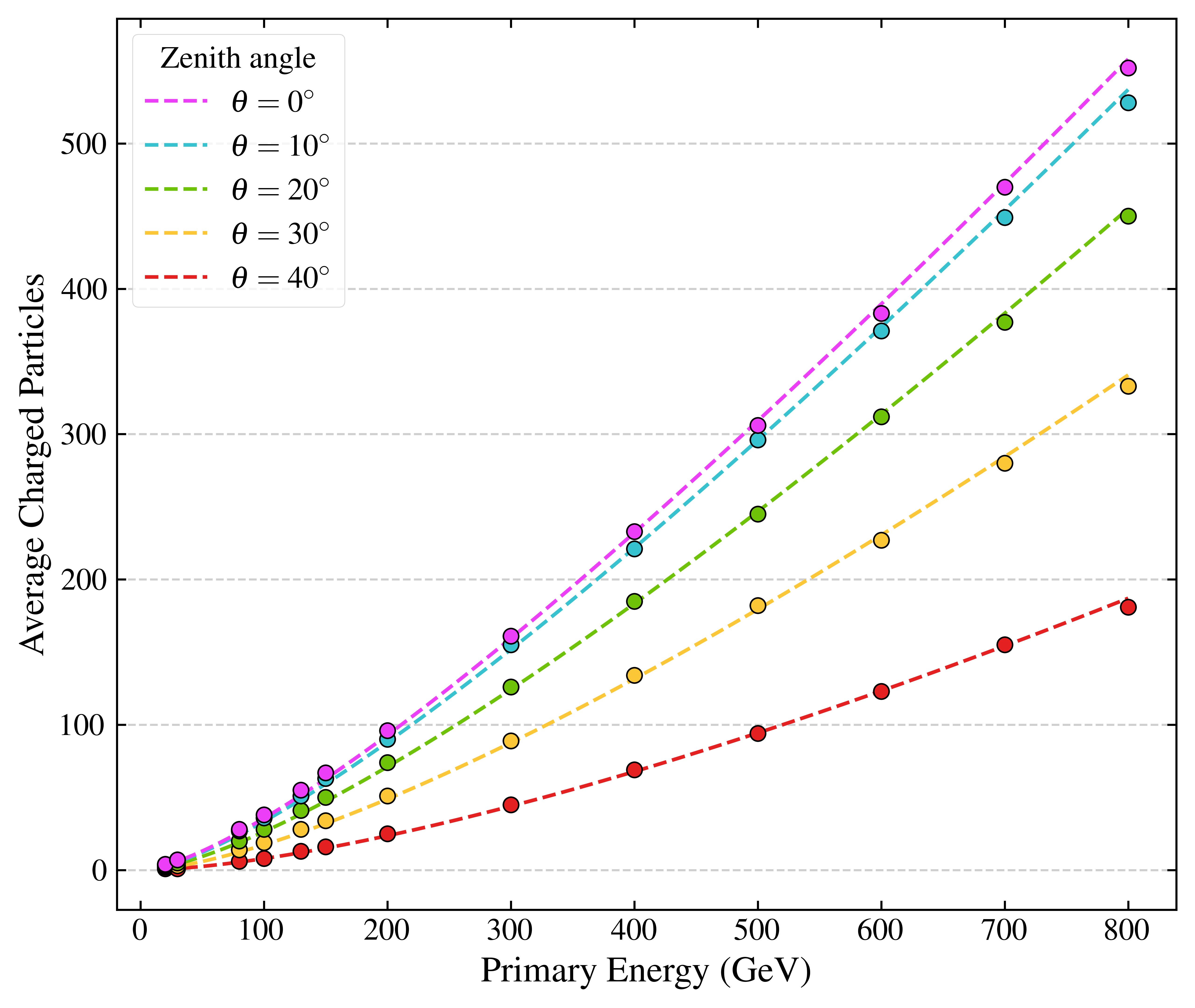}
    \caption{5900 m}
    \label{subfig:profile_1-5900}
\end{subfigure}
\caption{Shower profile fits of the Greisen profile to CORSIKA simulations at 5000 m, 5300 m, 5600 m and 5900 m.}
\label{fig:greisen_fits}
\end{figure}

Figure \ref{fig:greisen_fits} presents the fits of the Greisen profile to CORSIKA simulations at altitudes of 5000 m, 5300 m, 5600 m and 5900 m. Each subfigure plots the number of particles versus different energy of the gamma-ray, showing the Greisen profile closely tracking the CORSIKA simulation curves, particularly at the shower maximum where particle production peaks. As altitude increases from 5000 m to 5900 m, the shower maximum shifts to lower atmospheric depths due to the reduced air density, which decreases the radiation length and accelerates cascade development \cite{Weekes2002}. The modified profile’s ability to capture this shift is evident in the minimal residuals, especially in the pre-maximum and maximum regions, demonstrating its robustness for high-altitude conditions.

\begin{figure}[ht]
\centering
\begin{subfigure}[b]{0.47\textwidth}
    \centering
    \includegraphics[width=\textwidth]{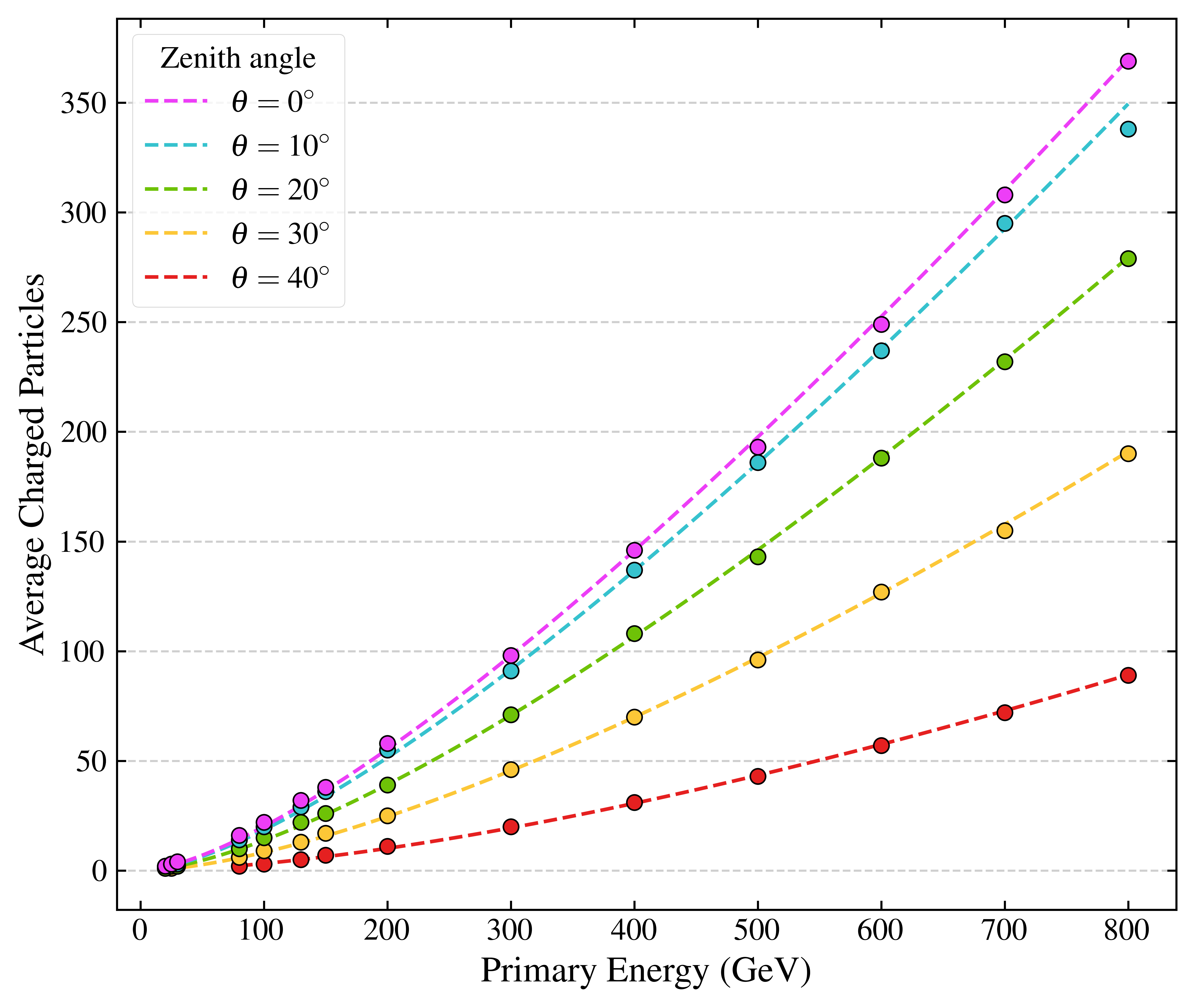}
    \caption{5000 m}
    \label{subfig:profile_1-5000}
\end{subfigure}
\hfill
\begin{subfigure}[b]{0.47\textwidth}
    \centering
    \includegraphics[width=\textwidth]{greisen_fit_5300m.png}
    \caption{5300 m}
    \label{subfig:profile_1-5300}
\end{subfigure}

\vspace{1em}

\begin{subfigure}[b]{0.47\textwidth}
    \centering
    \includegraphics[width=\textwidth]{greisen_fit_5600m.png}
    \caption{5600 m}
    \label{subfig:profile_1-5600}
\end{subfigure}
\hfill
\begin{subfigure}[b]{0.47\textwidth}
    \centering
    \includegraphics[width=\textwidth]{greisen_fit_5900m.png}
    \caption{5900 m}
    \label{subfig:profile_1-5900}
\end{subfigure}
\caption{Shower profiles fits of modifed Gresien profile to CORSIKA simulations at  5000 m, 5300 m, 5600 m and 5900 m.}
\label{fig:greisen_fits_alpha}
\end{figure}

\begin{figure}[ht]
\centering
\includegraphics[width=\columnwidth]{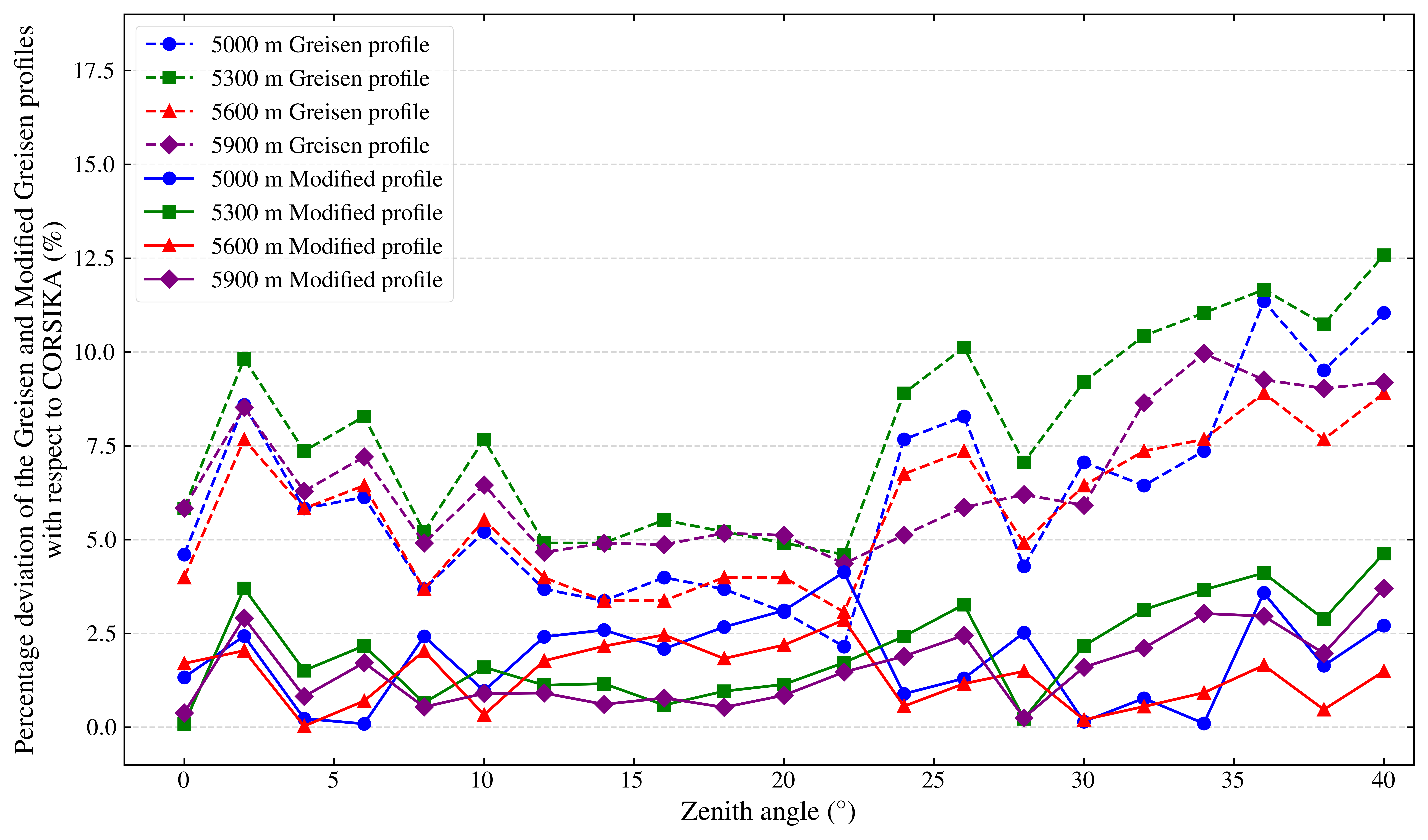}
\caption{Percentage deviation of fitted \(\alpha\) values from the classical and modified Greisen profiles at altitudes of 5000 m, 5300 m, 5600 m and 5900 m, as a function of zenith angle, demonstrating the modified profile’s reduced deviations.}
\label{fig:alpha_deviation}
\end{figure}

To quantify deviations between the classical and modified Greisen profiles and CORSIKA simulations, both profiles were fitted to the simulated particle numbers using a parameter \(\alpha\) to measure the quality of the fit, where \(\alpha = 1\) would indicate a perfect fit. The percentage deviation is defined as:
\begin{equation}
\text{Deviation (\%)} = \left|\alpha - 1\right| \times 100,
\label{eq:deviation}
\end{equation}
providing a direct metric for the results shown in Figure \ref{fig:alpha_deviation}.

providing a direct measure of the fit quality. Figure \ref{fig:alpha_deviation} shows these deviations across zenith angles and altitudes. The modified Greisen profile consistently achieves deviations below 4.7\%, while the classical profile reaches deviations up to 12.5\%, particularly at higher zenith angles (e.g., 40$^\circ$), where the increased path length (Equation \ref{eq:slant_depth}) amplifies scattering and absorption effects \cite{AlvarezMuniz2012}.

Table \ref{tab:greisen_fit} reports the fitted $\alpha$ values and the corresponding $\chi^2$ per degree of freedom ($\chi^2/\text{n.d.f.}$) for the classical Greisen profile across zenith angles (0$^\circ$--40$^\circ$) and altitudes (5000 m, 5300 m, 5600 m, and 5900 m). The \(\alpha\) values deviate from unity by up to 12.5\% (e.g., \(\alpha = 1.1258\) at 5300 m, 40$^\circ$), indicating systematic overestimation of particle numbers, particularly at larger zenith angles where path-length effects are significant. The \(\chi^2\)/n.d.f. values, ranging from 0.02 to 0.16, suggest reasonable fits but highlight limitations in the classical profile’s ability to model low-energy showers. Table \ref{tab:greisen_fit} presents the corresponding results for the modified Greisen profile, with \(\alpha\) deviations below 4.7\% (e.g., \(\alpha = 1.0463\) at 5300 m, 40$^\circ$) and consistently lower \(\chi^2\)/n.d.f. values, confirming the modified profile’s improved accuracy.

\section{Discussion}
\label{sec:discussion}

The results presented in Section \ref{sec:results} demonstrate that the modified Greisen profile, with its adjusted shower age term (Equation \ref{eq:modified_greisen_fct}), significantly outperforms the classical Greisen profile in aligning with CORSIKA simulations for low-energy cosmic gamma-ray showers (20--800 GeV). The modified profile achieves deviations in particle numbers below 4.7\%, as evidenced by the fitted \(\alpha\) values in Tables \ref{tab:greisen_fit} and, compared to deviations up to 12.5\% for the classical profile. Figure \ref{fig:alpha_deviation} illustrates this improvement, showing that the modified profile maintains consistently lower deviations (Equation \ref{eq:deviation}) across zenith angles, particularly at higher angles (e.g., 40$^\circ$) where the classical profile’s inaccuracies are exacerbated by increased atmospheric path lengths (Equation \ref{eq:slant_depth}). The low \(\chi^2\)/n.d.f. values, ranging from 0.02 to 0.16 across both profiles, indicate statistically robust fits, with the modified profile’s values closer to 1, reflecting a better match to the physical processes modeled by CORSIKA \cite{Gaisser1990}.

The performance of the modified Greisen profile stems from its empirical correction factor, which adjusts the shower age term to account for accelerated particle absorption in the post-maximum phase. At low energies (20--800 GeV), ionization losses dominate over radiative processes below the critical energy (\( E_{\text{cut}} \simeq 87 \) MeV), leading to faster particle attenuation than predicted by the classical profile’s high-energy approximations \cite{Lipari2009}. The modified profile mitigates this ensuring more accurate predictions of particle numbers in the decaying phase.

At high altitudes (5000--5900 m), the reduced air density (Equation \ref{eq:density}) accelerates shower development, shifting the shower maximum to lower atmospheric depths, as observed in Figure \ref{fig:greisen_fits}. Zenith-angle dependencies further complicate modeling, as oblique showers traverse a longer atmospheric path (Equation \ref{eq:slant_depth}), enhancing multiple Coulomb scattering and photon absorption \cite{Gaisser1990, AlvarezMuniz2012}. The modified profile’s adjustment better handles these effects, particularly at 5900 m, where the extended altitude tests the model’s scalability.

The importance of these findings lies in their implications for gamma-ray astrophysics at high-altitude observatories \cite{Abeysekara2017, Arratia2023}. The modified profile’s accuracy enables precise reconstruction of shower parameters, supporting studies of cosmic accelerators and dark matter searches. The computational efficiency of the analytical model, compared to the resource-intensive CORSIKA simulations, makes it a practical tool for real-time analysis, enhancing detection efficiency and supporting studies of cosmic gamma-ray sources and their role in galactic and extragalactic processes \cite{Weekes2002}. The reduced deviations (Figure \ref{fig:alpha_deviation}) enhance confidence in energy spectra and flux measurements, critical for astrophysical observations.

Despite these advances, limitations remain. The modified Greisen profile assumes a simplified atmospheric model (Equation \ref{eq:density}), neglecting seasonal variations in density, temperature, or humidity, which could affect shower development, particularly at extreme altitudes. CORSIKA’s reliance on predefined interaction cross-sections introduces potential systematic uncertainties, especially at the lowest energies (e.g., 20 GeV), where quantum electrodynamics approximations may be less accurate \cite{Lipari2009}. 
\section{Conclusion}  
\label{sec:conclusion}  

The modified Greisen profile demonstrates clear advantages over the classical formulation for modeling low-energy cosmic gamma-ray showers in the 20--800 GeV range, as validated against CORSIKA Monte Carlo simulations. Incorporating a zenith-angle--dependent correction factor, the modified profile achieves deviations in particle numbers below 4.7\% across altitudes of 5000--5900 m and zenith angles of 0$^\circ$--40$^\circ$, substantially improving upon the classical profile, which exhibits deviations up to 12.5\% (Tables \ref{tab:greisen_fit} and \ref{tab:modified_greisen_extended}, Figure \ref{fig:alpha_deviation}).  

Fitting procedures based on $\chi^2$ minimization confirm the statistical reliability of the modified profile, with low $\chi^2$/n.d.f. values (0.02--0.16) and tight alignment to CORSIKA simulations, particularly near the shower maximum (Figures \ref{fig:greisen_fits} and \ref{fig:greisen_fits_alpha}). By explicitly accounting for low-energy ionization losses and zenith-angle effects, the modified profile overcomes limitations inherent to the classical formulation, providing a physically consistent description of shower development under high-altitude conditions.  

The combination of accuracy, robustness, and computational efficiency establishes the modified Greisen profile as a practical tool for high-altitude observatories such as HAWC and the Condor Array. Its validated performance across a broad range of energies, zenith angles, and altitudes enhances capabilities for energy reconstruction, detection efficiency, and the discrimination of gamma-ray signals from hadronic backgrounds.  

These results position the modified Greisen profile as a reliable analytical framework for gamma-ray astrophysics. Future work should explore its extension to higher energies, the inclusion of dynamic atmospheric models and validation against observational datasets to further expand its applicability in the study of cosmic accelerators and related phenomena.

\begin{acknowledgments}
This work was funded by ANID PIA/APOYO AFB230003, Proyectos Internos de Investigación Multidisciplinarios USM 2024 \texttt{PI\_M\_24\_02} and Millennium Institute of Subatomic Physics at High Energy Frontier ICN2019\_044. The authors thank the High Performance Computing
Cluster (HPCC) at DFIS, UTFSM, San Joaquín, Santiago, Chile.
\end{acknowledgments}

\newpage
\appendix
\section{Appendix}
\label{sec:appendix}

\begin{table}[htbp]
\centering
\caption{Fitted \(\alpha\) and \(\chi^2\)/n.d.f. values from the \textbf{classical Greisen profile} for various zenith angles and altitudes}
\label{tab:greisen_fit}
\small
\begin{tabular}{c|cc|cc|cc|cc}
\toprule
\textbf{Angle (°)} & \multicolumn{2}{c|}{\textbf{5000 m}} & \multicolumn{2}{c|}{\textbf{5300 m}} & \multicolumn{2}{c|}{\textbf{5600 m}} & \multicolumn{2}{c}{\textbf{5900 m}} \\
                   & \(\alpha\) & \(\chi^2\)/n.d.f. & \(\alpha\) & \(\chi^2\)/n.d.f. & \(\alpha\) & \(\chi^2\)/n.d.f. & \(\alpha\) & \(\chi^2\)/n.d.f. \\
\midrule
0  & 1.0460 & 0.10 & 1.0583 & 0.12 & 1.0399 & 0.11 & 1.0584 & 0.12 \\
2  & 1.0859 & 0.11 & 1.0982 & 0.15 & 1.0767 & 0.16 & 1.0852 & 0.22 \\
4  & 1.0583 & 0.12 & 1.0736 & 0.16 & 1.0583 & 0.12 & 1.0629 & 0.19 \\
6  & 1.0613 & 0.08 & 1.0828 & 0.12 & 1.0644 & 0.16 & 1.0721 & 0.17 \\
8  & 1.0368 & 0.11 & 1.0521 & 0.09 & 1.0368 & 0.06 & 1.0491 & 0.15 \\
10 & 1.0521 & 0.11 & 1.0767 & 0.10 & 1.0552 & 0.14 & 1.0646 & 0.14 \\
12 & 1.0368 & 0.11 & 1.0491 & 0.08 & 1.0399 & 0.12 & 1.0466 & 0.16 \\
14 & 1.0337 & 0.07 & 1.0491 & 0.08 & 1.0337 & 0.10 & 1.0491 & 0.12 \\
16 & 1.0399 & 0.07 & 1.0552 & 0.10 & 1.0337 & 0.11 & 1.0487 & 0.13 \\
18 & 1.0368 & 0.11 & 1.0521 & 0.09 & 1.0399 & 0.10 & 1.0517 & 0.08 \\
20 & 1.0307 & 0.03 & 1.0491 & 0.07 & 1.0399 & 0.05 & 1.0512 & 0.09 \\
22 & 1.0215 & 0.04 & 1.0460 & 0.08 & 1.0307 & 0.08 & 1.0436 & 0.15 \\
24 & 1.0767 & 0.05 & 1.0890 & 0.11 & 1.0675 & 0.12 & 1.0512 & 0.15 \\
26 & 1.0828 & 0.08 & 1.1012 & 0.13 & 1.0736 & 0.12 & 1.0586 & 0.12 \\
28 & 1.0429 & 0.07 & 1.0706 & 0.07 & 1.0491 & 0.14 & 1.0620 & 0.13 \\
30 & 1.0706 & 0.07 & 1.0920 & 0.07 & 1.0644 & 0.09 & 1.0592 & 0.12 \\
32 & 1.0644 & 0.06 & 1.1043 & 0.11 & 1.0736 & 0.09 & 1.0865 & 0.14 \\
34 & 1.0736 & 0.03 & 1.1104 & 0.06 & 1.0767 & 0.07 & 1.0996 & 0.07 \\
36 & 1.1135 & 0.04 & 1.1166 & 0.08 & 1.0890 & 0.11 & 1.0926 & 0.11 \\
38 & 1.0951 & 0.04 & 1.1074 & 0.04 & 1.0767 & 0.05 & 1.0903 & 0.10 \\
40 & 1.1104 & 0.02 & 1.1258 & 0.10 & 1.0890 & 0.09 & 1.0919 & 0.06 \\
\bottomrule
\end{tabular}
\end{table}

\begin{table}[htbp]
\centering
\caption{Fitted \(\alpha\) and \(\chi^2\)/n.d.f. values from the \textbf{modified Greisen profile} for various zenith angles and altitudes.}
\label{tab:modified_greisen_extended}
\small
\begin{tabular}{c|cc|cc|cc|cc}
\toprule
\textbf{Angle (°)} & \multicolumn{2}{c|}{\textbf{5000 m}} & \multicolumn{2}{c|}{\textbf{5300 m}} & \multicolumn{2}{c|}{\textbf{5600 m}} & \multicolumn{2}{c}{\textbf{5900 m}} \\
                   & \(\alpha\) & \(\chi^2\)/n.d.f. & \(\alpha\) & \(\chi^2\)/n.d.f. & \(\alpha\) & \(\chi^2\)/n.d.f. & \(\alpha\) & \(\chi^2\)/n.d.f. \\
\midrule
0  & 0.9867 & 0.10 & 1.0008 & 0.12 & 0.9830 & 0.11 & 1.0038 & 0.12 \\
2  & 1.0243 & 0.11 & 1.0370 & 0.15 & 1.0204 & 0.16 & 1.0291 & 0.22 \\
4  & 0.9977 & 0.12 & 1.0151 & 0.16 & 1.0003 & 0.12 & 1.0082 & 0.19 \\
6  & 0.9991 & 0.08 & 1.0217 & 0.12 & 1.0070 & 0.16 & 1.0172 & 0.17 \\
8  & 0.9758 & 0.11 & 0.9935 & 0.09 & 0.9797 & 0.06 & 0.9946 & 0.15 \\
10 & 0.9903 & 0.11 & 1.0160 & 0.10 & 0.9967 & 0.14 & 1.0090 & 0.14 \\
12 & 0.9759 & 0.11 & 0.9888 & 0.08 & 0.9823 & 0.12 & 0.9909 & 0.16 \\
14 & 0.9741 & 0.07 & 0.9884 & 0.08 & 0.9784 & 0.10 & 0.9939 & 0.12 \\
16 & 0.9791 & 0.07 & 0.9941 & 0.10 & 0.9754 & 0.11 & 0.9922 & 0.13 \\
18 & 0.9733 & 0.11 & 0.9904 & 0.09 & 0.9817 & 0.10 & 0.9947 & 0.08 \\
20 & 0.9689 & 0.03 & 0.9886 & 0.07 & 0.9781 & 0.05 & 0.9915 & 0.09 \\
22 & 0.9587 & 0.04 & 0.9828 & 0.08 & 0.9714 & 0.08 & 0.9853 & 0.15 \\
24 & 1.0089 & 0.05 & 1.0242 & 0.11 & 1.0056 & 0.12 & 1.0189 & 0.15 \\
26 & 1.0130 & 0.08 & 1.0327 & 0.13 & 1.0116 & 0.12 & 1.0245 & 0.12 \\
28 & 0.9748 & 0.07 & 1.0023 & 0.07 & 0.9851 & 0.14 & 0.9975 & 0.13 \\
30 & 1.0015 & 0.07 & 1.0217 & 0.07 & 0.9980 & 0.09 & 1.0160 & 0.12 \\
32 & 0.9923 & 0.06 & 1.0313 & 0.11 & 1.0055 & 0.09 & 1.0211 & 0.14 \\
34 & 0.9990 & 0.03 & 1.0366 & 0.06 & 1.0092 & 0.07 & 1.0303 & 0.07 \\
36 & 1.0358 & 0.04 & 1.0411 & 0.08 & 1.0165 & 0.11 & 1.0296 & 0.11 \\
38 & 1.0164 & 0.04 & 1.0288 & 0.04 & 1.0047 & 0.05 & 1.0197 & 0.10 \\
40 & 1.0271 & 0.02 & 1.0463 & 0.10 & 1.0149 & 0.09 & 1.0370 & 0.06 \\
\bottomrule
\end{tabular}
\end{table}

\clearpage

\bibliographystyle{unsrt}
\bibliography{biblio}

\end{document}